\documentclass[aps,prl,preprint,amsmath,amssymb,showpacs,showkeys,floatfix]{revtex4-1}
\usepackage{graphicx}
\usepackage{tabulary}
\usepackage{array}
\usepackage{color}
\usepackage{bm}
\begin{document}

\date{\today}

\title{Effect of the shell material and confinement type on the conversion efficiency of the core/shell quantum dot nanocrystal solar cells}

\author{Mehmet Sahin}
\email{mehmet.sahin@agu.edu.tr; mehsahin@gmail.com} 
\affiliation{Abdullah G\"{u}l University, School of Engineering, Department of Material Science and Nanotechnology Engineering, S\"{u}mer Campus 38080 Kayseri, Turkey}

\begin{abstract}

In this study, effects of the shell material and confinement type on the conversion efficiency of the core/shell quantum dot nanocrystal (QDNC) solar cells have been investigated in a detail manner. For this purpose, the conventional, i.e original, detailed balance model, developed by Shockley and Queisser to calculate an upper limit for conversion efficiency of silicon p-n junction solar cells, is modified in a simple and an effective way and calculated the conversion efficiency of core/shell QDNC solar cells. Since the existing model relies on the gap energy ($E_g$) of the solar cell, it does not make an estimation about the effect of QDNC materials on the efficiency of the solar cells and gives the same efficiency values for several QDNC solar cells with the same $E_g$. The proposed modification, however, estimates a conversion efficiency in relation to the material properties and also confinement type of the QDNCs. The results of the modified model show that, in contrast to the original one, the conversion efficiencies of different QDNC solar cells, even if they have the same $E_g$, become different depending upon the confinement type and shell material of the core/shell QDNCs and this is crucial in design and fabrication of the new generation solar cells to predict the confinement type and also appropriate QDNC materials for better efficiency. 

\end{abstract}

\maketitle

\section{Introduction}

The last developments in the technology and wet chemical synthesizing techniques open a new door to fabricate of the new generation quantum dot nanocrystal (QDNC) based solar cells. It is expected that the new generation QDNC solar cells will have higher conversion efficiency\cite{noz1,noz2,sem,car,chu} since their electronic and optical properties are easily controlled through their size, structure and material composition and also the multiple exciton generation (MEG) can be possible in QDNCs. As well known, the QDNCs are nanoscale crystals of semiconductor materials in which the carriers can be completely confined in all spatial dimensions. This kind of confinement is gained some superiority, such as controlling of the effective band gap, to the QDNCs for some device applications. This and controllable these kinds of unique properties of QDNCs make them a good candidate to fabricate of new generation optoelectronic or photovoltaic devices~\cite{noz1,gre,akt,bro}.

The upper limit of conversion efficiency of a single p-n junction silicon solar cell calculated by detailed balance theory is approximately 33\% and this efficiency value is known as Shockley-Queisser limit~\cite{sho,que}. This limit has become a strong motivation for scientists working on development of solar cells and so, this model is very important in history of solar based energy studies. Although the model was reported in 1961 by Shockley and Queisser\cite{sho} for single p-n junction solar cells, it is used extensively as well to calculate the efficiency values of QDNC solar cells~\cite{leo,tom}. This model, fundamentally established depending on the band gap ($E_g$) variation, has been modified\cite{bre} in different manner to calculate the efficiency of new generation QDNC solar cells~\cite{han}. These important modifications are basically related to the MEG in QDNCs~\cite{bea,bin,smi}. In addition to the MEG, some other modifications such as, free carrier absorption, Auger recombination etc. have been realized by some authors~\cite{tli,tie}. The essential aim of these modifications is to execute a more realistic efficiency calculations and to understand the fundamental physics of the devices and as a result, to suggest much better QDNC solar cell designs. In some studies, the carrier multiplication phenomena and in addition to this, photon up- and down-conversion processes have been investigated in a detail manner to understand the limitation of high conversion efficiency of QDNC solar cells~\cite{wer,kli,tak,alh1,alh2,shp}.

As well known, the detailed balance model assumes that all photons coming from the sun with energies equal to or greater than $E_g$ are absorbed and formed electron-hole pairs (excitons). In this idealized model, the sole loss mechanism is radiative recombinations of the excitons~\cite{sho,tom}. Since the model is based on $E_g$ only, the efficiency of any photovoltaic device is equal to that of another one with the same $E_g$. For example, the efficiency of a QDNC solar cell with $E_g=1.1$ eV is almost completely same with that of a bulk silicon solar cell. Similarly, according to the original detailed-balance model, the efficiency of solar cells with type-I QDNC is identical with the efficiency of solar cells with the type-II QDNC if their $E_g$ values are the same~\cite{bin,smi}. There are a number of theoretical studies have been reported in the literature related to conversion efficiency of the QDNC based solar cells and the calculations have been performed in the frame of the original detailed balance model in all these studies~\cite{leo,tom,bin,smi,smi2,tli}.

Nevertheless, essentially, it is not possible to have the same efficiency values for all types of QDNC solar cells, even if they have the same $E_g$ owing to other material properties of solar cells such as, effective masses of the carriers, dielectric properties of the materials, size of the QDNCs, confinement type etc. and all these properties are very important in terms of the carrier dynamics in the solar cells. On the other hand, the recombination probability is so high in type-I structures and also, in practice, collecting of the carriers from type-I QDNCs is not so easy due to the both electron and hole confined inside the core when compared to the type-II QDNCs.

The main aim of this study is to modify the original detailed balance model in order to calculate the structure dependent upper limit for conversion efficiency and to investigate the effects of shell materials and confinement types of the QDNCs on the efficiency of the solar cells using the modified model. With this modification, the model can estimate an upper limit for the conversion efficiency of QDNC solar cells based on material properties and confinement type of the QDNCs. In the modification, without making drastic changes on the original model, the quantum mechanical oscillator strength effect is taken into account in the radiative recombination current calculations. As well known, the oscillator strength is an important and unitless parameter in determining of all optical properties of quantum mechanical systems from atoms to solids. The radiative recombination phenomenon in photovoltaic devices is also an optical process and the oscillator strength must be taken into account in the conversion efficiency calculations. As will see ahead, the oscillator strength is basically dependent on overlaps of the wavefunctions of the electron and hole as well as transition energy of the exciton and Kane energy of the materials. All these quantities rely on the crystal structure properties, confinement regime, effective masses of the carriers, and dielectric properties of the QDNC materials. In the next step, the conversion efficiencies of the solar cells based on type-I and type-II QDNC with different shell materials are calculated by using both the original and modified detailed balance models. The results are presented comparatively and probable physical reasons are discussed. We see that the modified model can estimate appropriate materials and confinement type of the QDNCs that will be used in design and fabrication of more efficient solar cells. 

\section{Model and Theory}
In the original detailed balance model, the photogenerated current density is given by~\cite{bre}

\begin{equation}
\label{eq:jpg}
J_{pg}=q_e \int_{E_g}^{\infty}QY(h\nu,E_g)\phi(h\nu)d(h\nu),
\end{equation}
where $q_e$ is the electronic charge, $\phi(h\nu)$ is the photon flux density of the sun\cite{sol}, and $QY(h\nu,E_g)$ is the quantum yield of the absorbed photon, dependent on both photon energy $h\nu$ and gap energy $E_g$. The $QY(h\nu,E_g)$ is actually external quantum efficiency ($EQE$) and it contains $EQE(h\nu)=C(h\nu)(1-R(h\nu))a(h\nu)$ and where $C(h\nu)$ is the collection probability of the excited carriers to do work, $R(h\nu)$ is reflectance of the incident photons and $a(h\nu)$ is the absorbance of incident photons. In ideal conditions, there is no reflectance, i.e. $R(h\nu)=0$, and all photons with equal or higher energies than $E_g$ are absorbed, i.e. $a(h\nu)=1$, and hence $C(h\nu)$ becomes equal to $QY(h\nu,E_g)$.

The MEG is integrated into the detailed balance model by favour of the $QY(h\nu,E_g)$ as

\begin{equation}
\label{eq:qy}
QY(h\nu,E_g)=\sum_{m=1}^M \theta(h\nu,mE_g).
\end{equation}
Here, $\theta(h\nu,mE_g)$ is Heaviside step function and $M$ is an integer, $M=\frac{h\nu_{max}} {E_g}$. In case of fixing of the $QY(h\nu,E_g)$ to unity, the MEG will become absent. The recombination current density in the original model is given by

\begin{equation}
\label{eq:rc}
J_{rc}=\frac{2\pi q_e}{h^3 c^2}\int_{E_g}^{\infty} \frac{QY(h\nu,E_g)(h\nu)^2}{e^{(h\nu-q_eVQY(h\nu,E_g))/kT}-1} d(h\nu),
\end{equation}
where $h$ is Planck's constant, $c$ is the light speed in the vacuum, $k$ is Boltzmann's constant, $T$ is temperature and the $V$ is applied voltage to the cell, and it is also taken into consideration as a constant quasi-Fermi level separation and its value is determined by a numerical search as it will maximize the efficiency of the solar cell. Here, the relation between the $QY(h\nu,E_g)$ and $EQE$ is apparently the same with that in the photogenerated current except absorbance parameter. For the recombination current density, the $EQE$ is taken as 

\begin{equation}
\label{eq:eqe}
EQE(h\nu)=C(h\nu)(1-R(h\nu))\epsilon(h\nu)
\end{equation}
where $\epsilon(h\nu)$ is the emissivity. That is, the absorbance parameter is replaced by the emissivity and its value is unity for the black body. The efficiency of the solar cell is calculated by means of

\begin{equation}
\label{eq:eff}
\eta=\frac{J_{net}V}{P_{in}},
\end{equation}
where $J_{net}=J_{pg}-J_{rc}$, net current density and $P_{in}$ is the total solar irradiance coming onto the solar cell and its value has been set to AM1.5 condition in the calculations.

\begin{figure}
    \centering
    \includegraphics[width=2.5in]{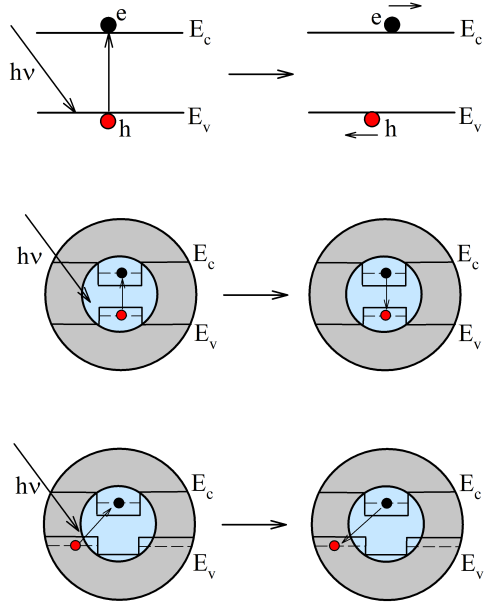}
    \caption{Schematic representation of the exciton forming in bulk semiconductors (top panel), in type-I QDNCs (middle panel), and in type-II QDNCs (bottom panel).}
    \label{fig:exc}
\end{figure}

In bulk semiconductor materials, as seen in the top panel of Fig.~\ref{fig:exc}, when an electron, with assistance of a photon, passes to the conduction band, it leaves a hole in the valence band and an attractive Coulomb potential comes into being between the electron and hole. On the other hand, the attractive Coulomb energy can not be predominant and it is small when compared to thermal energy ($kT$) for most bulk semiconductors and therefore, it can be easily broken down even if there is no an external electric field and hence, the carriers can move freely in the bands. That is, the recombination probability of an electron-hole couple is relatively weaker in bulk semiconductor materials.

As for QDNCs, there can be different recombination mechanisms depending on the confinement regimes. In type-I confinement regime, seen in middle panel of Fig.~\ref{fig:exc}, unlike bulk materials, since there are confinement potentials in both the conduction and valance bands, the electron and hole can not move like free particles. Therefore, the recombination probability of an exciton in type-I structures can be very high depending on size of the nanocrystal and the confinement potential depth even if there is an external electric field. Also, the attractive Coulomb potential between the electron-hole couple becomes predominant because of the confinement when compared to the bulk semiconductors. In type-II QDNCs shown in bottom panel of Fig.~\ref{fig:exc}, while one of the carriers is confined in the core, the other one is confined in the shell region and so the carriers are separated spatially in contrast to the type-I confinement regime. Consequently, it can be said that the recombination probability in type-II structures is smaller than that in type-I ones and therefore, the probability of contribution of the carriers to the photocurrent in type-II structures will be higher.

As the electron and hole move like free particles in bulk semiconductors, the recombination current mechanisms expressed by Eq.~\ref{eq:rc} in the original detailed balance model works good enough for solar cells like silicon p-n junction. On the other hand, in the QDNC solar cells, the recombination current density is larger because of higher recombination probabilities and hence, this higher recombination probability should be added into the recombination current density. Here, it should be noted that the oscillator strength will not be taken into account in calculation of the photogenerated current. Because, according to detailed balance limit assumptions, all photons with equal to or higher energies than $E_g$ are absorbed and collected to do work. On the other hand, the recombination current density is calculated statistically by basically using of Planck distribution function and it is strongly dependent on the radiative recombination oscillator strength. And hence, the oscillator strength must be taken into account in calculation of the recombination current density. Now, Eq.~\ref{eq:rc} can be modified as follows: The $EQE$ given in Eq.~\ref{eq:eqe} can be rearranged to include the recombination probability. When we focus on Eq.~\ref{eq:eqe}, in detailed balance limit and assumptions, we see that $R(h\nu)=0$, $\epsilon(h\nu)=1$ and $C(h\nu)=QY(h\nu,E_g)$. Here, the $C(h\nu)$ is strongly dependent on the overlap of electron-hole wavefunctions and this effect must be insert into Eq.~\ref{eq:eqe}. Therefore, it can be written as $C(h\nu)=fQY(h\nu,E_g)$ and in calculation of the recombination current density for a QDNC solar cell, employing of the following expression instead of Eq.~\ref{eq:rc} is more reasonable.

\begin{equation}
\label{eq:mrc}
J_{rc}=\frac{2\pi q_e}{h^3 c^2}\int_{E_g}^{\infty} \frac{f QY(h\nu,E_g)(h\nu)^2}{e^{(h\nu-fq_eVQY(h\nu,E_g))/kT}-1} d(h\nu),
\end{equation}
where $f$ is recombination oscillator strength and it is given by\cite{sah}

\begin{equation}
\label{eq:osc}
f=\frac{E_p}{2E_x} \left| \int \psi_e(r) \psi_h(r) d^3r \right|^2.
\end{equation}
Here, $E_p$ is Kane energy and $E_x$ is the exciton transition energy. $\psi_e(r)$ and $\psi_h(r)$ are the electron and hole wavefunctions. The energy states and the corresponding wavefunctions of the electron and hole are determined by solving of the Poisson-Schr\"odinger equations self-consistently. In calculations, excitonic effects (i.e Coulomb interactions between electron and hole) on both energy states and wavefunctions have been taken into account. All details of the electronic and optical properties calculations can be found in Refs.~\onlinecite{sah,koc}.

In Eq.~\ref{eq:osc}, the Kane energy is strongly dependent on the crystal properties of QDNC materials, and the wavefunctions involve the penetration effect to the barrier region and so, the electron and hole energy states are affected from the penetrations. That is, when we make an overall glance to the last two equations, we see that the recombination current density includes these material dependent parameters and it has been transformed into a material dependent form.

\section{Results and Discussion}

\begin{figure}
    \centering
    \includegraphics[width=3.5in]{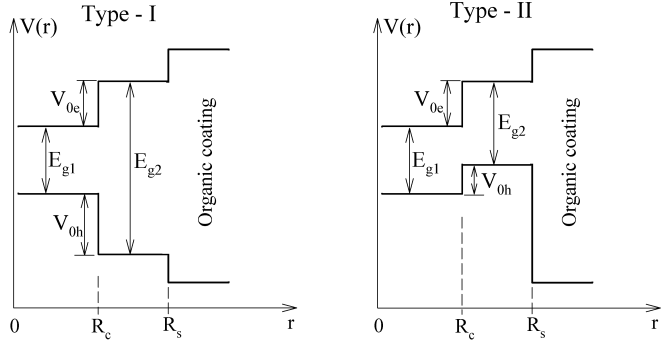}
    \caption{Schematic representation of the confinement potential profiles of the type-I and type-II QDNCs, respectively.}
    \label{fig:pot}
\end{figure}

In this study, both type-I, InP/ZnS and InP/GaP, and type-II, InP/GaAs and InP/GaSb, QDNC structures have been taken into consideration. In all structures, the core material is chosen as InP while the shell materials are different. Therefore, not only confinement types but also  effects of the shell materials can be evaluated more salutary. In addition, results of the modification on the original detailed balance model can be seen more clearly. The potential profiles of both type-I and type-II structures are seen in Fig.~\ref{fig:pot}. The band offsets have been determined by using of the electron affinity values of the materials.\cite{sah} The confinement potentials in the conduction bands, $V_{0e}$, are 0.60 eV and 0.75 eV for InP/ZnS and InP/GaP QDNCs, respectively, while the valance band confinements, $V_{0h}$, are 1.73 eV and 0.16 eV, respectively. In type-II structures, the conduction band offsets are 0.35 eV and 0.20 eV, and the valance band offsets are 0.27 eV and 0.92 eV for InP/GaAs and InP/GaSb QDNCs, respectively. In the calculations, the shell thickness is taken as constant, 10 \AA. All material parameters used in the electronic structure calculations are listed in Table~\ref{tab:table1}.

\begin{table*}
    \caption{\label{tab:table1}The material parameters used in the calculations.}
    \begin{ruledtabular}
        \begin{tabular}{ccccccc}
            Material & $m_e^*/m_0$  & $m_h^*/m_0$  & $\kappa$ & $E_g$ (eV) & $E_p$ (eV) &$\chi$ (eV)\\
            \hline
            InP   & 0.08\cite{ada}  & 0.69\cite{ada} & 12.9\cite{ada} & 1.35\cite{ada} & 17.0\cite{vur} & -4.50\cite{ada} \\
            ZnS   & 0.25\cite{ada}  & 0.59\cite{ada} & 8.9\cite{ada}  & 3.68\cite{ada} & 20.4\cite{van} & -3.90\cite{ada} \\
            GaP   & 0.114\cite{ada} & 0.52\cite{ada} & 11.0\cite{ada} & 2.26\cite{ada} & 28.0\cite{vur} &-3.75\cite{ada} \\
            GaAs  & 0.067\cite{ada} & 0.55\cite{ada} & 12.9\cite{ada} & 1.43\cite{ada} & 27.0\cite{vur} & -4.15\cite{ada} \\
            GaSb  & 0.039\cite{ada} & 0.37\cite{ada} & 15.5\cite{ada} & 0.72\cite{ada} & 27.0\cite{vur} & -4.21\cite{ada} \\
        \end{tabular}
    \end{ruledtabular}
\end{table*}

\begin{figure} 
    \centering
    \includegraphics[width=3in]{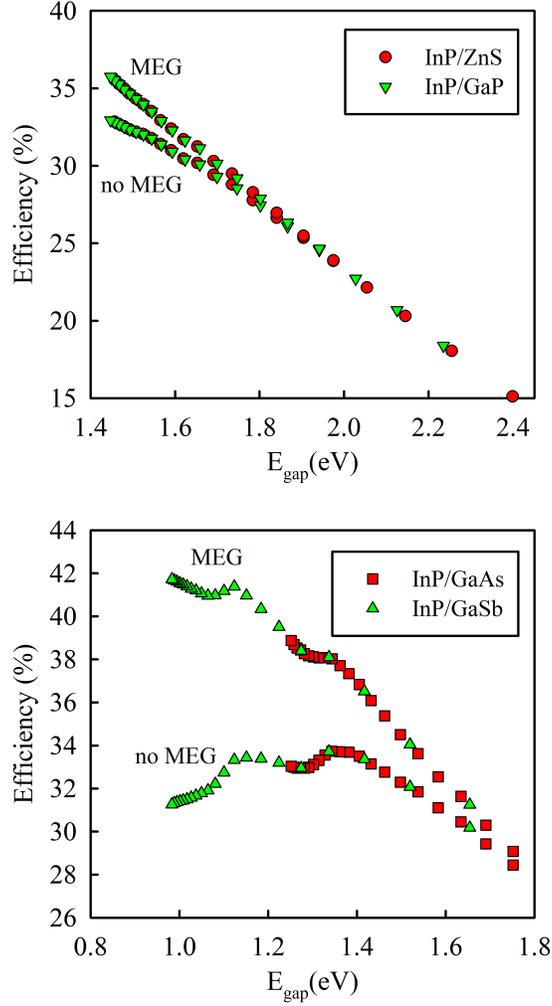}
    \caption{Efficiency values of QDNC solar cells for type-I, InP/ZnS and InP/GaP (top panel) and type-II, InP/GaAs and InP/GaSb (bottom panel). The results are obtained by original detailed-balance model.}
    \label{fig:eff_con}
\end{figure}

After computation of the energy levels and corresponding wavefunctions of the QDNCs, the photovoltaic efficiencies are calculated by using the original and modified detailed balance models and the results are plotted as a function of the $E_g$ of the QDNCs. The results of the original detailed-balance model are given in Fig.~\ref{fig:eff_con}. Top panel of the figure shows the efficiency values for type-I, InP/ZnS and InP/GaP, and the bottom panel demonstrates the efficiency values of type-II, InP/GaAs and InP/GaSb, QDNC solar cells. As seen from the top panel, in larger core radii which correspond to smaller $E_g$, the efficiency values of type-I QDNC solar cells are very high and decrease with increasing $E_g$ values of the QDNCs. Here, it is important to emphasize that there is no effect of what the shell material is on the solar cell efficiency because the original detailed balance model depends only on the $E_g$. When the MEG is considered in the calculations, the efficiency values become larger in case of $h\nu\geq 2E_g$ as expected and reported in previous studies~\cite{bin,smi}. It should be noted that the $QY(h\nu,E_g)$ is taken as maximum 2 when the MEG taken into consideration in the calculations because there can be maximum two carriers in ground states of the QDNCs for selected materials. When we look at bottom panel of Fig.~\ref{fig:eff_con}, we see that the maximum efficiency values are slightly greater with respect to the type-I structures because of the smaller $E_g$ values of the type-II QDNCs. Also the general behaviour of the efficiency values are the same with studies reported in the literature. As can be seen from the figure, the shell material has no effect on the solar cell efficiency in type-II structures as well. After this overall glances, when we focus on both panels of Fig.~\ref{fig:eff_con}, we see that the efficiency values are almost same in both type-I and type-II QDNC solar cells for the same $E_g$ values. In addition, the results are the same of p-n junction solar cell efficiency values reported in the literature. We can conclude that since the original detailed balance model gives an upper limit for efficiency values of solar cells depending on $E_g$ only, it does not give any information about the effects of shell materials and/or type of the QDNCs on the efficiencies of QDNC solar cell.

\begin{figure} [t!]
    \centering
    \includegraphics[width=3in]{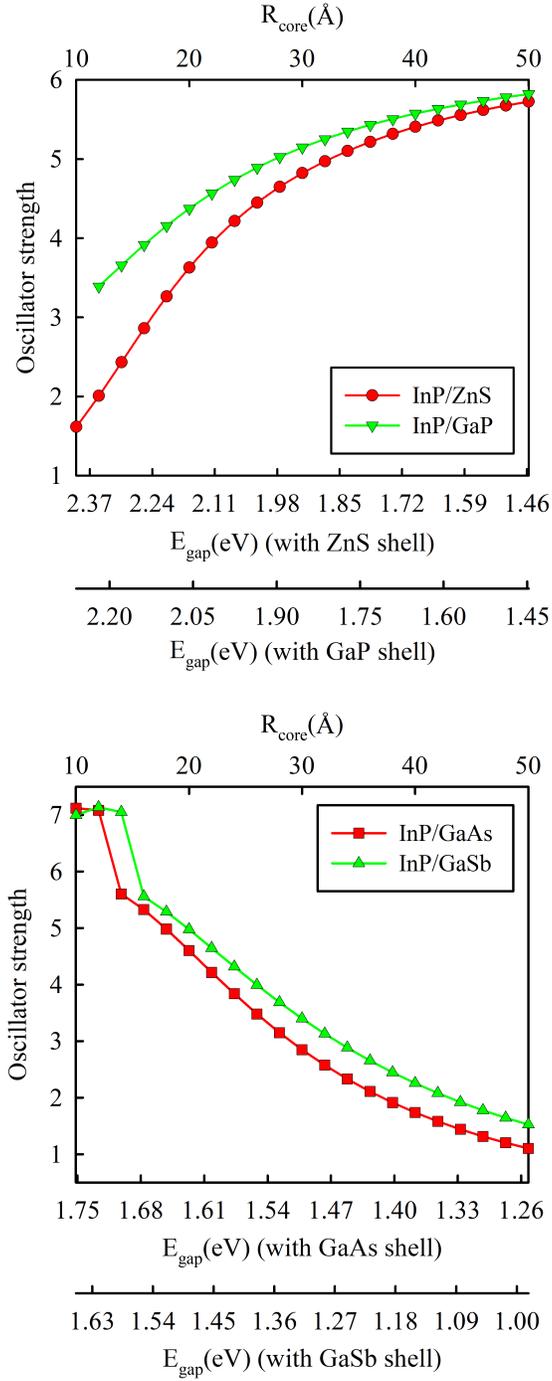}
    \caption{Oscillator strength of type-I, InP/ZnS and InP/GaP (top panel) and type-II, InP/GaAs and InP/GaSb (bottom panel) QDNCs.}
    \label{fig:osc}
\end{figure}

However, actually, the shell material and/or the confinement type of the QDNC must have crucial effects on the electronic and optical properties of QDNCs such as, overlaps of the wavefunctions, recombination oscillator strengths etc. depending on penetration of the wavefunctions to the shell regions as well as the other material parameters such as, effective masses, dielectric constants etc. Thereby, the recombination current density and so the efficiency of the QDNC based solar cells must be strongly dependent on these properties of the QDNCs. The oscillator strength (OS) contains all these effects and it must be taken into consideration in the recombination phenomenon as mentioned before. Figure~\ref{fig:osc} shows the OSs of the type-I (top panel) and type-II (bottom panel) QDNCs as a function of the core radii. Here, at the same time, the gap energies, corresponding to each core radius, are given on the bottom axes of the graphs. It should be noted that the gap energies are calculated by means of $E_g=E_g(bulk)+E_e+E_h$, where $E_g(bulk)=E_{g1}$ for the type-I, and $E_g(bulk)=E_{g1}-V_{0h}$ for the type-II structures, $E_e$ and $E_h$ are single particle energy states of the electron and hole, respectively. These single particle energy values are strongly dependent on the effective masses of the electron and hole and penetration of the wavefunctions to the barrier regions as mentioned in previous section and hence, the same core radii may correspond to different gap energies depending on the shell materials. When we look at both panels, we see that the behaviours of the OSs are completely different in type-I and type-II structures. The OS in type-I QDNCs increases with increasing core radius (decreasing $E_g$) while it decreases in type-II QDNCs. At the same time, the shell materials effects on the OS are seen clearly in all QDNCs. In type-I structures, the OS values of InP/ZnS are smaller, especially at smaller core radius, when compared to that of InP/GaP QDNC and their values increase with increasing core radius and become close to each other at larger core radii. This is because the overlaps of the electron and hole wavefunctions are larger in larger core radii. This results in higher recombination probabilities. Also, the overlapping is bigger in InP/GaP QDNC than in InP/ZnS. As for the type-II QDNCs, the OS values are almost same and bigger at smaller core radii (higher $E_g$ values) and decrease with increasing core radii. This is because both electron and hole localize to the vicinity of the core region at smaller core radii. When the core radius increases, strong confinement regime relaxes and the overlapping of the wavefunctions decreases with increasing spatial separation of the carriers. So the lifetime of the carriers becomes longer and the recombination probability becomes smaller.

\begin{figure} [t!]
    \centering
    \includegraphics[width=3in]{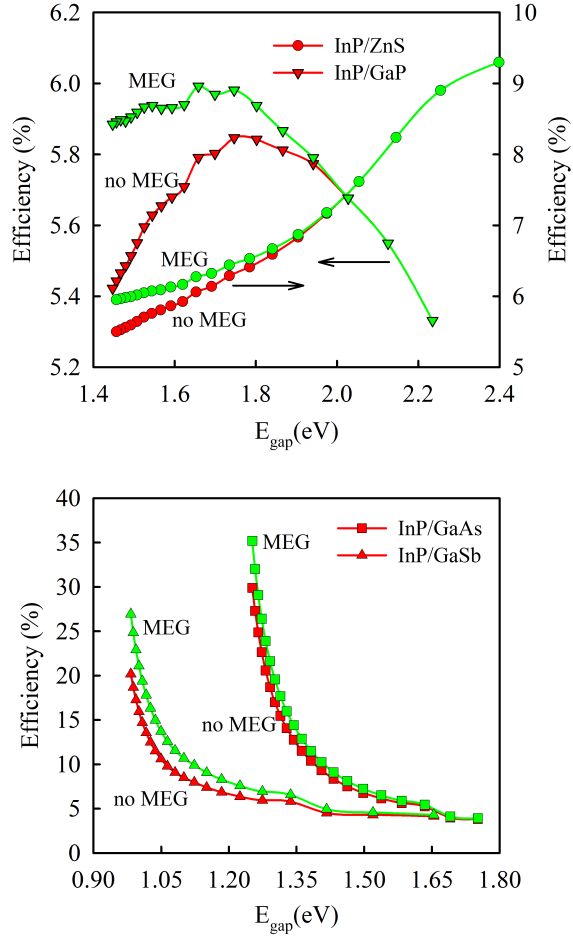}
    \caption{Efficiency values of QDNC solar cells for type-I, InP/ZnS and InP/GaP (top panel) and type-II, InP/GaAs and InP/GaSb (bottom panel). The results are obtained by modified detailed-balance model.}
    \label{fig:eff_mod}
\end{figure}

Figure~\ref{fig:eff_mod} shows the efficiency values calculated by the modified detailed balance model for type-I, InP/ZnS and InP/GaP, QDNC solar cells (top panel) and type-II, InP/GaAs and InP/GaSb, QDNC solar cells. In calculations, the recombination current densities are computed by means of Eq.~\ref{eq:mrc}. When we look at the both panels, we can see easily that the efficiency values are smaller than the results of original detailed balance calculations demonstrated in Fig.~\ref{fig:eff_con}. This situation is very drastic and at the same time more reasonable especially for type-I QDNC solar cells. If we focus on the efficiency values of InP/ZnS QDNC solar cell, we see a completely different behaviour. Here, the efficieny of InP/ZnS QDNC solar cell has smaller values for smaller band gap and increses with increasing $E_g$ and with further increasing of the gap value, they reach to maximum values which is slightly greater than $9\%$. The tendency of the efficiency values is not agreement with the result of the original model and exhibits a completely opposite character to it. This can be explained with smaller OS values at larger $E_g$ energies. On other hands, although changes in the efficiency values of the InP/GaP QDNC solar cells with the $E_g$ are similar to original method results reported in the literature, the values are smaller drastically than the original method results. This is also because of the higher OS values. Since the OS of InP/GaP QDNC is larger than that of InP/ZnS QDNC, the efficiency values of InP/ZnS QDNC solar cell are higher. The efficiency of the InP/GaP QDNC solar cell reaches to maximum $6\%$ values. As can be seen from top panel of the figure, although the core materials of the QDNCs are the same, the efficiency values exhibit completely different behaviour depending on the shell materials. Many electronic and optical properties of core/shell QDNCs such as, oscillator strength are affected from the shell material as expected and mentioned above. The effect of the oscillator strength on the efficiency is seen clearly in type-I QDNC solar cells. The results of modified model is very reasonable because, until now, there is no an experimental study related to type-I core/shell QDNC solar cells reported with higher efficiency values more than 4\% ~\cite{kum}. Indeed, the type-I QDNCs are more compatible for LEDs rather than solar cells. It should be noted that there are many different reasons rather than OS that make suitable of type-I QDNCs for LED applications such as, confinement strength, dielectric properties of the QDNCs, defect states etc. The illumination intensity of InP/GaP LED's is more efficient than that of InP/ZnS ones~\cite{kim} and this indirectly shows that the efficiency of InP/GaP QDNC solar cells will be smaller than that of InP/ZnS QDNC solar cells.

If we look at the bottom panel of Fig.~\ref{fig:eff_mod}, although the maximum efficiency values are relatively in a good agreement with the literature~\cite{leo,tom,smi}, we see completely different behaviors in efficiency values with respect to the $E_g$ when compared to bottom panel of Fig.~\ref{fig:eff_con}. When we compare the bottom panels of Figs.~\ref{fig:eff_con} and ~\ref{fig:eff_mod}, we see that while the efficiency values of InP/GaSb QDNC solar cell is higher in the first figure, the efficiency values of InP/GaAs QDNC solar cell becomes robust in the second one. In Fig.~\ref{fig:eff_mod}, the efficiency values start from higher ones and decrease rapidly with increasing $E_g$ in both type-II structures as similar to reported studies. But here, the decreasing is continuously in contrast to bottom panel of Fig.~\ref{fig:eff_con} and reported studies in the literature\cite{bea,bin,smi}. In type-II structures as well, the effect of the oscillator strength can be seen explicitly. If we perform an overall evaluation, we see that the efficiency values are higher in type-II QDNC solar cells when compared to results of type-I. Also the behavioral characters of efficiency with the $E_g$ can be completely different in both type-I and type-II QDNC solar cells depending on the shell materials of the structures. As can be seen clearly from the results, the modified model gives an upper limit depending on the materials and confinement type for conversion efficiency of QDNC solar cells.

\section{Conclusion}
The original detailed balance model has been modified and the upper limits for conversion effeciency values of QDNC based solar cells have been calculated by using both the original and modified models for core/shell type-I and type-II QDNCs with different shell materials. The modification is executed on the recombination current density. The original detailed balance model gives the same upper limit values and the same tendencies for all QDNC solar cells with the same $E_g$. That is, the original model does not provide any information about the effects of materials and/or confinement type of the QDNCs on the solar cell conversion efficiencies. Indeed, both confinement type and shell material have enormous influence on the recombination rate and hence, this must affect the recombination current density. Since the modification is taken into consideration the recombination rate with assistance of the oscillator strength, the modified model yields different upper efficiency values for different QDNC solar cells even if they have the same $E_g$ values. That is, the modified model is able to estimate an upper limit for the efficiency of the QDNC solar cells depending on the material properties and confinement types. This is an extremely important result in terms of both confinement type and determining of the QDNC materials which will be fabricated for the solar cell applications. It is hoped that the modified model will be used to carry out more realistic efficiency calculations in the manner of including material properties and so better QDNC solar cell design can be realized.

\section*{Acknowledgments} 
The author thanks Abdullah Gul University Foundation (AGUV) for their partial financial support.

\end{document}